\documentclass[conference]{IEEEtran}
\IEEEoverridecommandlockouts

\usepackage[sort]{cite}
\usepackage{amsmath,amssymb,amsfonts}
\usepackage{amsthm}
 \usepackage[sort,compress]{natbib}
\usepackage{amsmath}

 \usepackage{hyperref}

\usepackage{pifont} 

\usepackage{enumitem}
\usepackage{subcaption}

\usepackage[linesnumbered,ruled,vlined]{algorithm2e}

\usepackage{graphicx}
\usepackage{diagbox}
\usepackage{lscape}
\usepackage{multicol}
\usepackage{textcomp}
\usepackage{tabularx}
 
\usepackage{booktabs}  
\usepackage{array}     
\usepackage{caption}   
\usepackage{multirow}  

\usepackage{xcolor}
\usepackage{listings}
\usepackage{etoolbox}
\usepackage{tikz}
\usepackage{bm}
\def\BibTeX{{\rm B\kern-.05em{\sc i\kern-.025em b}\kern-.08em
    T\kern-.1667em\lower.7ex\hbox{E}\kern-.125emX}}

\usepackage{etoolbox}
\makeatletter
\patchcmd{\@makecaption}
  {\scshape}
  {}
  {}
  {}
\makeatother

\lstdefinelanguage{CUDA}{
    morekeywords={float,void,double},
    morecomment=[l]{//},
    morecomment=[s]{/*}{*/},
    morestring=[b]"
}

\begin{document}

\title{FlashMP: Fast Discrete Transform-Based Solver for Preconditioning Maxwell’s Equations on GPUs
\thanks{This work is supported by the Strategic Priority Research Program of Chinese Academy of Sciences, Grant NO.XDB0500101.}\\
}
\author{
\IEEEauthorblockN{
Haoyuan Zhang\textsuperscript{1,2}, 
Yaqian
Gao\textsuperscript{1,2},
Xinxin
Zhang\textsuperscript{1,2}, 
Jialin
Li\textsuperscript{1,2},
Runfeng
Jin\textsuperscript{1,2},\\ 
Yidong
Chen\textsuperscript{3}, 
Feng
Zhang\textsuperscript{1,2},
Wu
Yuan\textsuperscript{1,2}, 
Wenpeng
Ma\textsuperscript{4},
Shan	Liang\textsuperscript{1,2}, 
Jian Zhang\textsuperscript{1,2} and 
Zhonghua Lu\textsuperscript{1,2}  
}
 
\IEEEauthorblockA{
\textsuperscript{1}Computer Network Information Center, Chinese Academy of Sciences, Beijing, China\\
\textsuperscript{2}University of Chinese Academy of Sciences, Beijing, China
\IEEEauthorblockA{
\textsuperscript{3}Tsinghua University, 
Beijing, China\\
\IEEEauthorblockA{\textsuperscript{4}Xinyang Normal University, Xinyang, China}
Email:  \{zhanghaoyuan, gaoyaqian, zhangxinxin, lijialin, jinrunfeng, zhangfeng, zhlu\}@cnic.cn,\\ \{yuanwu, liangshan, zhangjian\}@sccas.cn }  }
  chenyidong@tsinghua.edu.cn, mawp@xynu.edu.cn
}

\maketitle

\begin{abstract}
Efficiently solving large-scale linear systems is a critical challenge in electromagnetic simulations, particularly when using the Crank-Nicolson Finite-Difference Time-Domain method. Existing iterative solvers are commonly employed to handle the resulting sparse systems but suffer from slow convergence due to the ill-conditioned nature of the double-curl operator. Approximate preconditioners, like SOR and Incomplete LU decomposition (ILU) provide insufficient convergence, while direct solvers are impractical due to excessive memory requirements. To address this, we propose FlashMP, a novel preconditioning system that designs a subdomain exact solver based on discrete transforms.  FlashMP provides an efficient GPU implementation that achieves multi-GPU scalability through domain decomposition. Evaluations on AMD MI60 GPU clusters (up to 1000 GPUs) show that FlashMP reduces iteration counts by up to 16$\times$ and achieves speedups of 2.5$\times$ to 4.9$\times$ compared to baseline implementations in state-of-the-art libraries Hypre. Weak scalability tests show parallel efficiencies up to 84.1\%.
\end{abstract}


\section{Introduction}

Numerical simulation of electromagnetic phenomena governed by Maxwell's equations is vital in science and engineering, with applications in antenna design \cite{1330474}, radar systems \cite{PENG20142387}, photonic crystals \cite{Sun2021Transient}, and geophysical inversion \cite{shubitidze2018crank}. These fields demand accurate and efficient methods to model complex behaviors, but increasing needs for precision and scale challenge traditional approaches. As the complexity of electromagnetic systems grows, the demand for high-resolution simulations and detailed modeling of intricate geometries becomes more pronounced, pushing the boundaries of computational resources and numerical techniques.

The Crank-Nicolson Finite-Difference Time-Domain (CN-FDTD) method \cite{1330474,2015Improvement}, prized for its unconditional stability and energy conservation, is ideal for long-term simulations. These properties make it particularly suitable for applications requiring extended time horizons. However, large-scale CN-FDTD simulations require solving sparse linear systems from the double-curl operator \cite{PENG20142387,8543126} at each time step, a task demanding efficient solvers for high-resolution or complex geometries. The computational cost of solving these systems can become prohibitive, especially when dealing with fine discretizations or large-scale problems.

Current solutions use iterative methods like BiCGSTAB \cite{1330474,rouf2009solution,garcia2010efficient,xu2010gpu} and GMRES \cite{rouf2009solution,yang2008application,qiang2004cn}, as direct solvers are infeasible for large problems. While these iterative methods are widely adopted due to their lower memory footprint and adaptability to large-scale problems, they face significant challenges in achieving practical acceleration on GPU platforms. Specifically, existing preconditioning techniques, such as Jacobi, IC \cite{PENG20142387}, and ILU \cite{qiang2004cn,rouf2009solution}, often fail to provide substantial speedup on GPUs. Fast Transform-based Preconditioners (FTP) \cite{CHABORY20087755} are grid-size-independent but limited to a single subdomain, while direct solvers like LU \cite{wei2018domain,qiang2004cn} are impractical due to overhead.

Moreover, the parallel efficiency of large-scale simulations is severely constrained by communication overhead and poor adaptation to the spectral properties of Maxwell's equations \cite{CHABORY20087755,wei2018domain}. Existing methods often struggle to achieve high parallel efficiency on distributed GPU architectures, especially when scaling to thousands of GPUs. This limitation is further exacerbated by the lack of techniques that simultaneously reduce the number of iterations and the computational cost per iteration. As a result, there is a significant gap in the literature regarding practical preconditioning strategies that can effectively leverage the parallelism of GPUs while maintaining robustness and efficiency.

To address these challenges, we propose \textit{FlashMP}, a preconditioner for efficient CN-FDTD simulations on GPUs. \textit{FlashMP} employs discrete transforms via SVD to decouple the double-curl operator into \(3 \times 3\) systems, significantly reducing computational complexity by employing a subdomain exact solver as the preconditioner, \textit{FlashMP} minimizes the number of convergence steps and global communication overhead. Efficient mapping operators on GPUs ensure low computational cost per iteration. This marks the first time practical speedups have been achieved in large-scale electromagnetic simulations on GPU clusters.

This work offers three key contributions:
\begin{itemize}
    \item A novel subdomain exact solver design based on discrete transforms to decouple and solve subdomain problems for the double-curl operator efficiently.
    \item An efficient GPU implementation that leverages domain decomposition to achieve multi-GPU scalability.
    \item A comprehensive performance analysis, quantifying \textit{FlashMP}'s convergence, time breakdown, and scalability through experiments with BiCGSTAB and GMRES, demonstrating up to 16$\times$ reduction in iteration counts, 2.5$\times$ to 4.9$\times$ speedups, and 84.1\% parallel efficiency.
\end{itemize}

\section{Background} 

\subsection{Model Problem}

We consider the Maxwell’s curl equations for an isotropic media which can be written in the form:
{\small
\begin{equation}
\nabla \times E = -\frac{\partial B}{\partial t}, \quad \nabla \times H = \frac{\partial D}{\partial t}
\label{eqn:pde}
\end{equation}}\noindent
where \(E\), \(H\) are electric and magnetic fields, \(D = \epsilon E\), \(B = \mu H\), with \(\epsilon\), \(\mu\) as permittivity and permeability.

The Crank-Nicolson (CN) scheme solves the discretized Maxwell's equations by a full time step size with one marching procedure, and averages the right-hand-sides of the discretized Maxwell's equations at \(t+1\) time step and \(t\) time step:
{\small
\begin{equation}
\epsilon \frac{E^{t+1} - E^t}{\Delta t} = \frac{1}{2} \left[ \nabla \times H^{t+1} + \nabla \times H^t \right]
\label{eqn:E}
\end{equation}
\begin{equation}
\mu \frac{H^{t+1} - H^t}{\Delta t} = -\frac{1}{2} \left[ \nabla \times E^{t+1} + \nabla \times E^t \right]
\label{eqn:H}
\end{equation}}
Assuming \(\epsilon = \mu = 1\) (normalized units, common in simulations) and substituting \eqref{eqn:H} into \eqref{eqn:E}, we obtain:
{\small
\begin{equation}
E^{t+1} + \frac{\Delta t^2}{4} \nabla \times \nabla \times E^{t+1} = E^t + \Delta t \nabla \times H^t - \frac{\Delta t^2}{4} \nabla \times \nabla \times E^t
\label{eqn:E_final}
\end{equation}
The update equations for electric field components can be written in the following linear system form with a discrete double-curl operator:
{\small
\begin{equation}
\mathbf{A} E^{t+1} = R
\label{eqn:A}
\end{equation}}}\noindent
where \(\mathbf{A} = \mathbf{I} + \frac{\Delta t^2}{4} \nabla \times \nabla \times\), and \(R\) includes \(E^t\) and  \(H^t\). Only \(E^{t+1}\) is unknown, as \(H^{t+1}\) is computed via \eqref{eqn:H}.

\begin{table}[t]
\centering
\caption{Summary of Preconditioners for Solving Maxwell's Equations using CN-FDTD. ``GPU?'' means whether GPU acceleration is utilized. ``P.C.'' stands for preconditioner.}
\begin{tabular}{cccccc}
\toprule
\textbf{Ref.} & \textbf{Iterative Solver} & \textbf{P.C. Type} & \textbf{GPU?} & \textbf{Scale} \\
\midrule
\ \cite{1330474}         & \texttt{BiCGSTAB}         & ILU         & \textcolor{red}{\ding{55}} & 1 CPU \\
\ \cite{PENG20142387}    & \texttt{CG}                     & IC              & \textcolor{red}{\ding{55}} & 1 CPU \\
\ \cite{xu2010gpu}  & \texttt{BiCGSTAB}                   & NONE             & \textcolor{blue}{\ding{51}} & 1 GPU \\
\ \cite{yang2008application}        & \texttt{GMRES}  & SAI-SSOR &  \textcolor{red}{\ding{55}} & 1 CPU \\
\ \cite{wei2018domain}        & \texttt{NONE}                  & DD-LU            & \textcolor{red}{\ding{55}} & 4 CPUs \\
\ \textbf{Ours}                  & \textbf{\texttt{BiCGSTAB} / \texttt{GMRES}}           & \textbf{FlashMP}         & \textcolor{blue}{\ding{51}} & \textbf{1000 GPUs} \\
\bottomrule
\label{tab:previous_works}
\end{tabular}
\end{table}

\subsection{High Performance Linear Solvers}

Solving sparse linear systems \(\mathbf{A} E = R\) is central to computational electromagnetics, enabling applications like photonic crystals and unexploded ordnance detection \cite{1330474,8543126, Sun2021Transient, PENG20142387,rouf2009solution,shubitidze2018crank,qiang2004cn,garcia2010efficient,xu2010gpu}. These systems arise from discretizing Maxwell’s equations into meshes, a necessity since most PDEs lack analytical solutions. The Crank-Nicholson Finite-Difference Time-Domain (CN-FDTD) method, valued for its unconditional stability and energy conservation, avoids the Courant-Friedrichs-Lewy (CFL) constraint, producing sparse matrices with mostly zero entries for efficient storage.

Solvers are either direct or iterative. Direct methods like LU or QR factorization adapt dense matrix techniques but suffer from fill-in, increasing memory use and limiting scalability. Iterative solvers, such as BiCGStab \cite{1330474,rouf2009solution,garcia2010efficient,xu2010gpu}, GMRES \cite{rouf2009solution,yang2008application,qiang2004cn} are preferred for large systems but converge slowly with the ill-conditioned double-curl operator \cite{PENG20142387,rouf2009solution}, necessitating preconditioners.

\textbf{Preconditioning Iterative Methods.} Preconditioning accelerates iterative solvers like BiCGStab and GMRES by using a matrix \(\mathbf{M}\) that approximates \(\mathbf{A}\), transforming the system into \(\mathbf{M}^{-1}\mathbf{A}E = \mathbf{M}^{-1} R\) for faster convergence \cite{1330474,8543126, Sun2021Transient, PENG20142387,rouf2009solution,shubitidze2018crank,qiang2004cn,garcia2010efficient,xu2010gpu}. \(\mathbf{M}\) must be easier to invert than \(\mathbf{A}\), a challenge for ill-conditioned double-curl operators \cite{PENG20142387,rouf2009solution}. Approximate preconditioners like Jacobi and ICCG \cite{PENG20142387} are cost-effective but converge slowly. Fast Transform-based Preconditioners (FTP) \cite{CHABORY20087755} offer grid-size-independent iterations but are limited to a single subdomain. Two-step methods like SAI-SSOR \cite{yang2008application} or RCM \cite{wei2018domain} improve convergence but struggle with severe ill-conditioning. ILU preconditioners \cite{qiang2004cn,rouf2009solution} balance robustness and cost by controlling fill-in, while direct solvers like LU \cite{wei2018domain,qiang2004cn} reduce iterations but incur high overhead, highlighting a trade-off: better approximations of \(\mathbf{A}\) speed convergence at the cost of per-iteration complexity.

\textbf{Domain Decomposition.} To improve scalability, domain decomposition is a widely used technique that splits a matrix into subdomains lying along the diagonal, and each subdomain is solved \cite{wei2018domain, smith1997domain}. The subdomain can be performed using ILU for approximate solutions or LU for complete solutions, and the resultant vectors are combined to approximate the solution of the entire matrix. When subdomains are overlapped, Restricted Additive Schwarz (RAS) \cite{cai1999restricted} is needed to combine the subdomain solutions to improve the approximation. We use RAS as a global preconditioner, which is given by:
{\small
\begin{equation*}
\mathbf{M}_{RAS}^{-1} = \sum_{i=1}^p \mathbf{S}_i^{0^T} \mathbf{A}_i^{-1} \mathbf{S}_i^\gamma
\end{equation*}}
where \(\mathbf{S}_i^\gamma\) is the restriction operator associated with the \(i^{\text{th}}\) subdomain with a overlap-\(\gamma\), \(\mathbf{A}_i = \mathbf{S}_i^\gamma \mathbf{A} \mathbf{S}_i^{\gamma^T}\) is the restricted submatrix from \(\mathbf{A}\) to the \(i^{\text{th}}\) subdomain, and \(\mathbf{S}_i^0\) is the restriction operator associated with the \(i^{\text{th}}\) subdomain without overlap. The discrete transform-based exact subdomain solver proposed later in this paper efficiently computes \(\mathbf{A}_i^{-1}\), enabling fast and scalable preconditioning.

\begin{figure*}[h]
  \centering
  \includegraphics[width=1.00\textwidth]{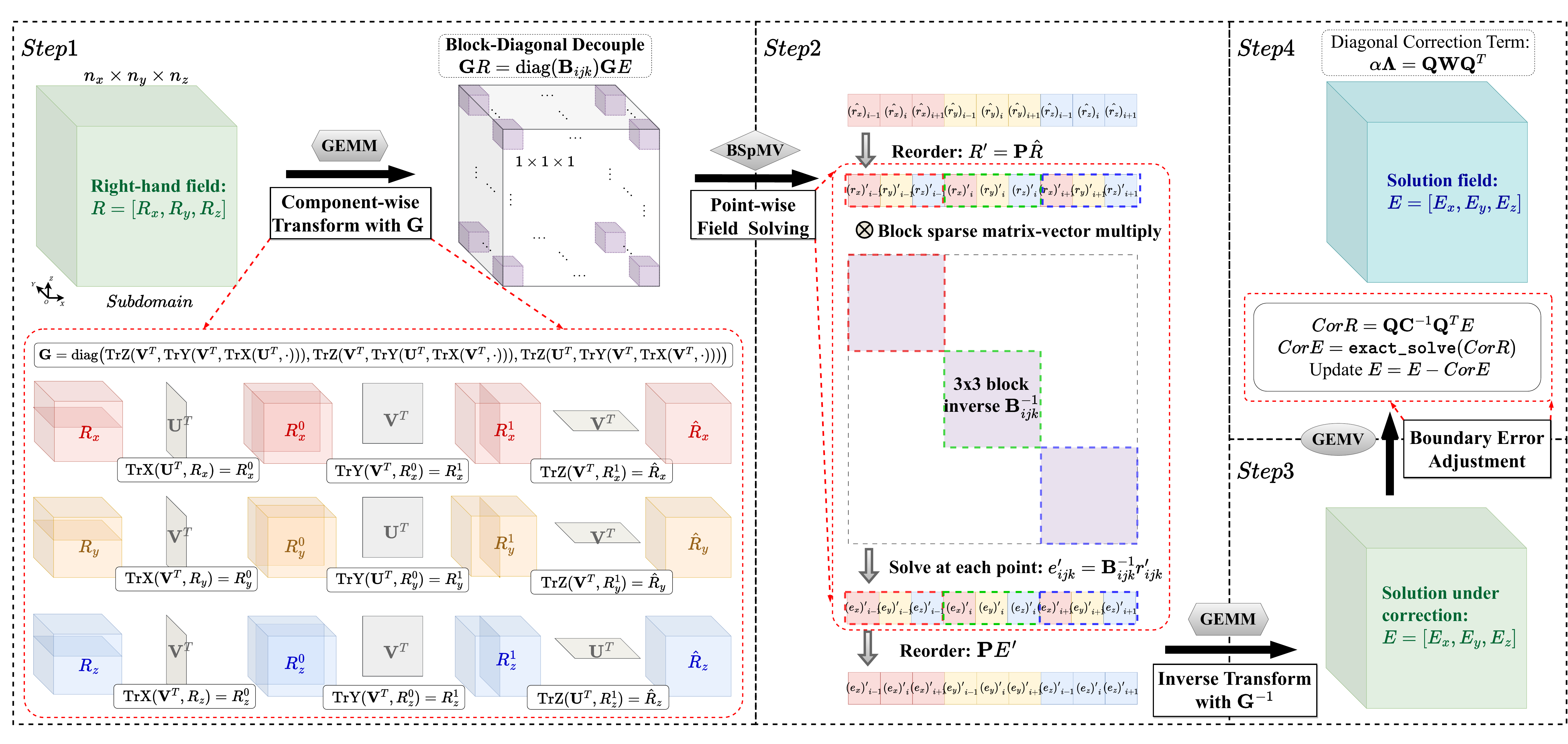}
  \caption{Steps of subdomain exact solving with discrete transform and low-rank correction.}
  \label{fig:overview}
\end{figure*}


\section{Innovation Implementation}

We introduce a preconditioning system, FlashMP, to efficiently speed up iterative solvers on multi-GPUs. For algorithmic optimization, we design a subdomain exact solver based on discrete transform (illustrated in Figure \ref{fig:overview}). In light of the following analysis, we can systematically derive Algorithm \ref{alg:subdomain_solving}.

\subsection{Algorithmic Innovations}

We consider a 3D grid with subdomain size \( n_x = n_y = n_z = n \). Our goal is to solve the linear system \(\mathbf{A} E = R\), where \(\mathbf{A} \in \mathbb{R}^{3n^3 \times 3n^3}\) is the coefficient matrix, \(E = [E_x, E_y, E_z]^T \in \mathbb{R}^{3n^3 \times 1}\) is the electric field vector with components in the \(x\), \(y\), and \(z\) directions, and \(R \in \mathbb{R}^{3n^3 \times 1}\) is the right-hand side vector, on a grid with \( n^3 \) points (\( 3n^3 \) variables). 

We note the double-curl operator using first-order forward (\(\mathbf{D}_l^f \in \mathbb{R}^{n^3 \times n^3}\)) and backward (\(\mathbf{D}_l^b = -(\mathbf{D}_l^f)^T\)) difference operators \citep{DifferenceOperatorsMatrices}, where \(l \in \{x, y, z\}\), then the curl operator is represented as:

\begin{equation}
\nabla \times E = \left[ \scriptsize
\begin{array}{c}
-\partial_z E_y + \partial_y E_z \\
\partial_z E_x - \partial_x E_x \\
-\partial_y E_x + \partial_x E_y
\end{array} \right] = \left[ \scriptsize
\begin{array}{ccc}
0 & -\mathbf{D}_z & \mathbf{D}_y \\
\mathbf{D}_z & 0 & -\mathbf{D}_x \\
-\mathbf{D}_y & \mathbf{D}_x & 0
\end{array} \right]  E
\end{equation}
where \(\mathbf{D}_x \in \{\mathbf{D}_x^f, \mathbf{D}_x^b\}\), \(\mathbf{D}_y \in \{\mathbf{D}_y^f, \mathbf{D}_y^b\}\), and \(\mathbf{D}_z \in \{\mathbf{D}_z^f, \mathbf{D}_z^b\}\) are the first-order difference operators in the \(x\), \(y\), and \(z\) directions, respectively. The double-curl operator in (\ref{eqn:A}) becomes:

{\small
\begin{equation*}
\mathbf{M} = \left[
\renewcommand{\arraystretch}{1.5} 
\scriptsize
\begin{array}{ccc}
-\mathbf{D}_z^b \mathbf{D}_z^f - \mathbf{D}_y^b \mathbf{D}_y^f & \mathbf{D}_y^b \mathbf{D}_x^f & \mathbf{D}_z^b \mathbf{D}_y^f \\
\mathbf{D}_x^b \mathbf{D}_y^f & -\mathbf{D}_z^b \mathbf{D}_z^f - \mathbf{D}_x^b \mathbf{D}_x^f & \mathbf{D}_z^b \mathbf{D}_y^f \\
\mathbf{D}_x^b \mathbf{D}_z^f & \mathbf{D}_y^b \mathbf{D}_z^f & -\mathbf{D}_y^b \mathbf{D}_y^f - \mathbf{D}_x^b \mathbf{D}_x^f
\end{array}
\right]
\end{equation*}
}

Define \(\alpha = \frac{\Delta t^2}{4}\), so the linear system in (\ref{eqn:A}) has the following representation:
{\small
\begin{equation}
(\mathbf{I} + \alpha \mathbf{M}) E = R
\label{eqn:linear_sys}
\end{equation}
}\noindent
where \(\mathbf{A} = \mathbf{I} + \alpha \mathbf{M}\), and \(\mathbf{I} \in \mathbb{R}^{3n^3 \times 3n^3}\) is the identity matrix.

\textbf{Observation.} The linear system (\ref{eqn:linear_sys}) arises from a non-symmetric double-curl operator \(\mathbf{M}\) with cross-derivative terms, precluding traditional diagonalization methods used for symmetric operators like the Laplacian \citep{inoue1992numerical}. Forward and backward difference operators exhibit symmetry (\(\mathbf{D}_l^b = -(\mathbf{D}_l^f)^T\)), suggesting shared spectral properties.

\textbf{Key Idea.} We can employ a discrete transform based on the SVD of the forward difference operator (\(\mathbf{D}^f = \mathbf{U} \mathbf{S} \mathbf{V}^T\)). The symmetry \(\mathbf{D}_l^b = -(\mathbf{D}_l^f)^T\) allows reuse of the SVD (\(\mathbf{D}_l^b = -\mathbf{V} \mathbf{S} \mathbf{U}^T\)). We aim to diagonalize \(\mathbf{A}\) to decouple variables across grid points for efficient solving.

We define the compact difference matrix \(\mathbf{D}^f \in \mathbb{R}^{n \times n}\) as:

{\footnotesize
\[
\mathbf{D}^f = \begin{pmatrix}
-1 & 1 & 0 & \cdots & 0 \\
0 & -1 & 1 & \cdots & 0 \\
0 & 0 & -1 & \cdots & 0 \\
\vdots & \vdots & \vdots & \ddots & \vdots \\
0 & 0 & 0 & \cdots & -1
\end{pmatrix}
\]}
Perform singular value decomposition (SVD) on \(\mathbf{D}^f\):
{\small
\begin{equation}
\mathbf{D}^f = \mathbf{U} \mathbf{S} \mathbf{V}^T
\label{eqn:svd}
\end{equation}
}\noindent
where \(\mathbf{S} = \text{diag}(\sigma_1, \dots, \sigma_n) \in \mathbb{R}^{n \times n}\), and \(\mathbf{U}, \mathbf{V} \in \mathbb{R}^{n \times n}\) are orthogonal matrices.

Let \(F\) denote any component of the electric field \(E\) along the \(x\), \(y\), or \(z\)-direction, i.e., \(F \in \{E_x, E_y, E_z\}\). Define tensor product operators \(\text{TrX}, \text{TrY}, \text{TrZ}\) for a 3D array \(F \in \mathbb{R}^{n^3 \times 1}\), reshaped from a tensor of size \(n \times n \times n\), which perform discrete transforms on the field.
{\small
\begin{equation}
\begin{aligned}
\text{TrX}(\mathbf{T}, F) &= \sum_{m=1}^n t_{im} f_{mjk} \\
\text{TrY}(\mathbf{T}, F) &= \sum_{m=1}^n t_{jm} f_{imk} \\
\text{TrZ}(\mathbf{T}, F) &= \sum_{m=1}^n t_{km} f_{ijm}
\end{aligned}
\label{eqn:tensor}
\end{equation}
}\noindent
where \(\mathbf{T} \in \mathbb{R}^{n \times n}\), and \(f_{ijk}\) represents the \((i,j,k)\)-th element of \(F\) in tensor form. Thus, it is evident that:
{\small
\begin{equation}
\begin{aligned}
\mathbf{D}_x^f F &= \text{TrX}(\mathbf{D}^f, F) \\
\mathbf{D}_y^f F &= \text{TrY}(\mathbf{D}^f, F) \\
\mathbf{D}_z^f F &= \text{TrZ}(\mathbf{D}^f, F)
\end{aligned}
\label{eqn:trxyz}
\end{equation}
}\noindent
Define the transform operator \(\mathbf{G} \in \mathbb{R}^{3n^3 \times 3n^3}\):

{\small
\begin{equation*}
\begin{aligned}
\mathbf{G} &= \text{diag}\Big( \text{TrZ}(\mathbf{V}^T, \text{TrY}(\mathbf{V}^T, \text{TrX}(\mathbf{U}^T, \cdot))) \\
&\phantom{= \text{diag}\Big\{} \text{TrZ}(\mathbf{V}^T, \text{TrY}(\mathbf{U}^T, \text{TrX}(\mathbf{V}^T, \cdot))) \\
&\phantom{= \text{diag}\Big\{} \text{TrZ}(\mathbf{U}^T, \text{TrY}(\mathbf{V}^T, \text{TrX}(\mathbf{V}^T, \cdot))) \Big)
\end{aligned}
\label{eqn:G}
\end{equation*}
}

The operator definition facilitates deriving the diagonalized form of \(\mathbf{A}\), followed by four steps for exact subdomain solve:

\textbf{Step 1: Component-wise transform.} We apply \(\mathbf{G}\) to transform the system (\ref{eqn:linear_sys}):
{\small
\(  
\mathbf{G} (\mathbf{I} + \alpha \mathbf{M}) \mathbf{G}^{-1} (\mathbf{G} E) = \mathbf{G} R
\) 
}\noindent
, yielding:
{\small
\begin{equation}
(\mathbf{I} + \alpha \mathbf{H}) \hat{E} = \hat{R}
\label{eqn:H_linear_sys}
\end{equation}
}\noindent
where \(\hat{E} = \mathbf{G} E\), \(\hat{R} = \mathbf{G} R\), and \(\mathbf{H} = \mathbf{G} \mathbf{M} \mathbf{G}^{-1} \in \mathbb{R}^{3n^3 \times 3n^3}\). By substituting (\ref{eqn:svd}) and (\ref{eqn:trxyz}) into (\ref{eqn:H_linear_sys}), we can obtain the form of \(\mathbf{H}\) as follows:

\begin{equation*}
\mathbf{H} = \scalebox{0.75}{$
\left[
\renewcommand{\arraystretch}{1.6} 
\scriptsize 
\begin{array}{ccc}
\text{TrZ}(\mathbf{S}^T \mathbf{S}, \cdot) + \text{TrY}(\mathbf{S}^T \mathbf{S}, \cdot) & -\text{TrY}(\mathbf{S}^T, \text{TrX}(\mathbf{S}, \cdot)) & -\text{TrZ}(\mathbf{S}^T, \text{TrX}(\mathbf{S}, \cdot)) \\
-\text{TrX}(\mathbf{S}^T, \text{TrY}(\mathbf{S}, \cdot)) & \text{TrZ}(\mathbf{S}^T \mathbf{S}, \cdot) + \text{TrX}(\mathbf{S}^T \mathbf{S}, \cdot) & -\text{TrZ}(\mathbf{S}^T, \text{TrY}(\mathbf{S}, \cdot)) \\
-\text{TrX}(\mathbf{S}^T, \text{TrZ}(\mathbf{S}, \cdot)) & -\text{TrY}(\mathbf{S}^T, \text{TrZ}(\mathbf{S}, \cdot)) & \text{TrX}(\mathbf{S}^T \mathbf{S}, \cdot) + \text{TrY}(\mathbf{S}^T \mathbf{S}, \cdot)
\end{array}
\right]
$}
\end{equation*}

Expanding the \(x\)-direction component of (\ref{eqn:H_linear_sys}):
{\small
\begin{equation*}
\begin{aligned}
\hat{R}_x &= \hat{E}_x + \alpha \text{TrZ}(\mathbf{S}^T \mathbf{S}, \hat{E}_x) + \alpha \text{TrY}(\mathbf{S}^T \mathbf{S}, \hat{E}_x) \\
&\quad + \alpha \text{TrY}(\mathbf{S}^T, \text{TrX}(\mathbf{S}, \hat{E}_y)) + \alpha \text{TrZ}(\mathbf{S}^T, \text{TrX}(\mathbf{S}, \hat{E}_z)) 
\end{aligned}
\end{equation*}
}\noindent

The scalar form of the above equation is as follows:
{\small
\begin{equation*}
\begin{aligned}
(\hat{r}_x)_{ijk} &= (1 + \alpha \sigma_k^2 + \alpha \sigma_j^2) (\hat{e}_x)_{ijk} + \alpha \sigma_i \sigma_j (\hat{e}_y)_{ijk} + \alpha \sigma_i \sigma_k (\hat{e}_z)_{ijk}
\end{aligned}
\end{equation*}
}\noindent
where \(\hat{e}_x, \hat{e}_y, \hat{e}_z\) and \(\hat{r}_x, \hat{r}_y, \hat{r}_z\) are the scalar elements at grid point \((i,j,k)\) of the solution field components \(\hat{E}_x, \hat{E}_y, \hat{E}_z\) and right-hand components \(\hat{R}_x, \hat{R}_y, \hat{R}_z\), respectively.

For the \(y\)- and \(z\)-directions we have:
{\small
\begin{equation*}
(\hat{r}_y)_{ijk} = -\alpha \sigma_i \sigma_j (\hat{e}_x)_{ijk} + (1 + \alpha \sigma_i^2 + \alpha \sigma_k^2) (\hat{e}_y)_{ijk} - \alpha \sigma_j \sigma_k (\hat{e}_z)_{ijk}
\end{equation*}
\begin{equation*}
(\hat{r}_z)_{ijk} = -\alpha \sigma_i \sigma_k (\hat{e}_x)_{ijk} - \alpha \sigma_j \sigma_k (\hat{e}_y)_{ijk} + (1 + \alpha \sigma_i^2 + \alpha \sigma_j^2) (\hat{e}_z)_{ijk}
\end{equation*}
}

Define the block matrix:
{\small
\begin{equation*}
\mathbf{B}_{ijk} = \mathbf{I} + \alpha \begin{bmatrix}
\sigma_j^2 + \sigma_k^2 & -\sigma_i \sigma_j & -\sigma_i \sigma_k \\
-\sigma_i \sigma_j & \sigma_i^2 + \sigma_k^2 & -\sigma_j \sigma_k \\
-\sigma_i \sigma_k & -\sigma_j \sigma_k & \sigma_i^2 + \sigma_j^2
\end{bmatrix} \in \mathbb{R}^{3 \times 3}
\end{equation*}
}\noindent
we can get:


{\small
\begin{equation}
\mathbf{B}_{ijk} \begin{bmatrix}
(\hat{e}_x)_{ijk} \,   (\hat{e}_y)_{ijk} \,   (\hat{e}_z)_{ijk}
\end{bmatrix}^T = \begin{bmatrix}
(\hat{r}_x)_{ijk} \,   (\hat{r}_y)_{ijk} \,  (\hat{r}_z)_{ijk}
\end{bmatrix}^T
\label{eqn:rbe}
\end{equation}
}\noindent
(\ref{eqn:rbe}) indicates that the linear system of size \(3n^3 \times 3n^3\) in (\ref{eqn:linear_sys}) can be decoupled into \(n^3\) small linear systems of size \(3 \times 3\).

\textbf{Step 2: Point-wise field solving.} We define the permutation matrix \(\mathbf{P} \in \mathbb{R}^{3n^3 \times 3n^3}\), which essentially reorders the indices \(l,i, j, k\) (where \(l \in \{x, y, z\}\) indicates the three components of field variable in \(x, y, z\)  respectively, and \(i, j, k\) represent the grid indices along the \(x, y, z\) directions).
{\small
\begin{equation*}
\begin{aligned}
\mathbf{P} R &= \mathbf{P} \left\{ \left\{ r_{x, ijk}  \right\}, \left\{ r_{y, ijk} \right\}, \left\{ r_{z, ijk}  \right\} \right\} \\
&= \left\{ \left\{ r_{x, ijk}, r_{y, ijk}, r_{z, ijk} \mid 1 \leq i,j,k \leq n \right\} \right\}
\end{aligned}
\end{equation*}
}
By reordering both sides of (\ref{eqn:H_linear_sys}), an equivalent equation is obtained:
{\small
\begin{equation*}
\text{diag} \left( \mathbf{B}_{ijk} \right) \mathbf{P}^T \hat{E} = \mathbf{P} \hat{R}
\label{qgn:PR}
\end{equation*}
}\noindent


\textbf{Step 3: Component-wise inverse transform.}
We can get \texttt{exact\_solve} process for solving (\ref{eqn:linear_sys}): (1) compute \(R' = \mathbf{P} \mathbf{G} R\), (2) solve \(3 \times 3\) system \(\mathbf{B}_{ijk} (E')_{ijk} = (R')_{ijk}\) at each grid point, (3) recover \(E = \mathbf{G}^{-1} \mathbf{P} E'\). Here, \(\mathbf{G}\) and \(\mathbf{G}^{-1}\) are used for the transform and inverse transform, respectively.

\begin{algorithm}
\caption{Subdomain exact solving with discrete transform and low-rank correction.}
\label{alg:subdomain_solving}
\KwIn{grid size \(n\), right-hand field \(R \in \mathbb{R}^{3n^3 \times 1}\), precomputed {\small\(\mathbf{U} \in \mathbb{R}^{n \times n}\), \(\mathbf{S} \in \mathbb{R}^{n \times 1}\), \(\mathbf{V}^T \in \mathbb{R}^{n \times n}\), \(\text{diag} \left( \mathbf{B}^{-1}_{ijk} \right)\),       \(\mathbf{C}^{-1} \in \mathbb{R}^{(6n^2-3n) \times (6n^2-3n)}\)}}
\KwOut{Solution field \(E = \mathbf{A}^{-1} R\)}
\(E \gets \texttt{exact\_solve}(R)\)\;
\tcc{Correct for boundary errors}
{
  \(\text{CorR} \gets \mathbf{Q} \mathbf{C}^{-1} \mathbf{Q}^T E\)\tcp*{\textcolor{blue}{GEMV}}
  \(\text{CorE} \gets \texttt{exact\_solve}(\text{CorR})\)\;
  \emph{Update} \(E \gets E - \text{CorE}\)\;
}
\KwRet{\(E\)}\;
\BlankLine
\SetKwFunction{FDirectSolve}{exact\_solve}
\SetKwProg{Fn}{Function}{:}{}
\Fn{\FDirectSolve{\(X\)}}{
  \KwIn{Vector \(X \in \mathbb{R}^{3n^3 \times 1}\)}
  \KwOut{Vector \(Y \in \mathbb{R}^{3n^3 \times 1}\)}
  Component-wise transform: \(\hat{X} \gets \mathbf{G} X\)\tcp*{\textcolor{blue}{GEMM}}
  Reorder \(\hat{X}\) into grid-major: \(X' \gets \mathbf{P} \hat{X}\)\;
  \tcc{Point-wise field solving}
   \((Y')_{ijk} \gets \mathbf{B}_{ijk}^{-1} (X')_{ijk}\) \tcp*{\textcolor{blue}{BSpMV}}
  Reorder \(Y'\) into component-major: \(\hat{Y} \gets \mathbf{P}^T Y'\)\;
  Inverse transform: \(Y \gets \mathbf{G}^{-1} \hat{Y}\)\tcp*{\textcolor{blue}{GEMM}}
  \KwRet{\(Y\)}
}
\end{algorithm}

\begin{figure*}[h]
  \centering
  \includegraphics[width=1.0\textwidth]{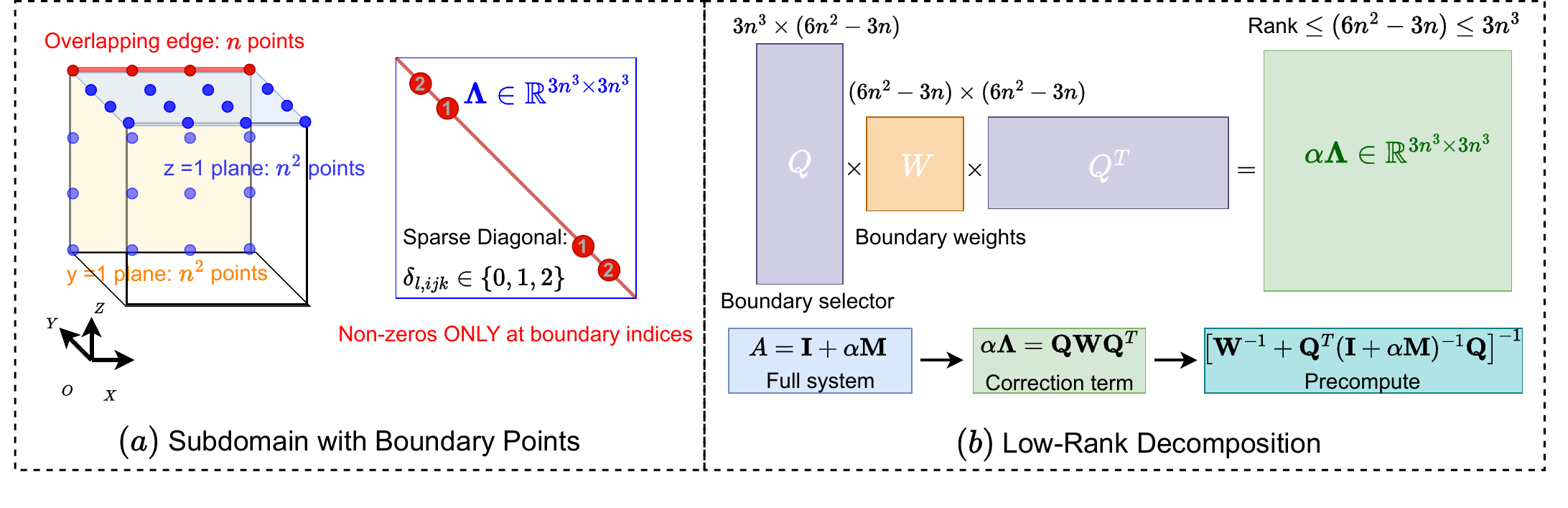}
  \caption{Illustration of Low-Rank Boundary Correction. 
(a) Left: Why low-rank? Only surfaces matter, not volume. Total non-zeros: \(2n^2 - n\) for the \(x\) component. (a) Right: \(\mathbf{\Lambda}\)'s sparse diagonal with non-zeros (e.g., 2 at edges, 1 at faces) at boundary indices. 
(b) Low-rank decomposition \(\mathbf{Q} \mathbf{W} \mathbf{Q}^T\), from a tall-skinny \(\mathbf{Q}\) to a small \(\mathbf{W}\), yielding a rank \(\leq 6n^2 - 3n \ll 3n^3\), followed by the application of the Woodbury formula.}
  \label{fig:low_rank}
\end{figure*}

\textbf{Step 4: Boundary error correction.}
The system in (\ref{eqn:linear_sys}) assumes Dirichlet boundary conditions \(E|_{\partial \Omega} = 0\). However, the difference operators \(\mathbf{D}_l^f\) and \(\mathbf{D}_l^b\) near the boundary points do not explicitly enforce these constraints in the transformed system, leading to boundary errors that must be corrected to ensure exact subdomain solving. To address this, we introduce a sparse diagonal correction term \(\alpha \mathbf{\Lambda}\), which encodes the boundary effects, resulting in the corrected linear system:
\begin{equation}
(\mathbf{I} + \alpha \mathbf{M} + \alpha \mathbf{\Lambda}) E = R,
\label{eqn:error_linear_sys}
\end{equation}
where \(\mathbf{\Lambda} = \text{diag}(\delta_{l,ijk} \mid l=x,y,z, 1 \leq i,j,k \leq n) \in \mathbb{R}^{3n^3 \times 3n^3}\), \(E, R \in \mathbb{R}^{3n^3 \times 1}\), and \(\delta_{l,ijk} \in \{0, 1, 2\}\) encodes the boundary effects. For direction \(l \in \{x, y, z\}\), define the indices \(p_l\) and \(q_l\) as:
\[
(p_l, q_l) = \begin{cases}
(j, k) & \text{if } l = x \\
(i, k) & \text{if } l = y \\
(i, j) & \text{if } l = z
\end{cases}
\]
Then, the function \(\delta_l(l, i, j, k)\) is given by:
\begin{equation*}
\delta_l(l, i, j, k) = \begin{cases}
2 & \text{if } p_l = q_l = 1 \\
1 & \text{if exactly one of } p_l = 1 \text{ or } q_l = 1 \\
0 & \text{otherwise}
\end{cases}
\end{equation*}

The sparse pattern of \(\mathbf{\Lambda}\) is characterized by non-zero entries confined to boundary points of each field component, reflecting the geometric insight that errors arise from asymmetric differencing at boundaries and propagate along surfaces rather than through the volume. As visualized in Figure \ref{fig:low_rank}(a), the 3D subdomain (left) with total non-zeros \(2n^2 - n\) for the \(x\) component (two planes of \(n^2\) points each minus \(n\) overlapping edge points); across three components, this yields \(6n^2 - 3n\). The matrix \(\mathbf{\Lambda}\) (right) shows non-zeros (e.g., 2 at edges, 1 at faces) only along boundary indices in an otherwise zero-filled diagonal. This surface-like scaling (\(O(n^2)\)) versus volumetric (\(O(n^3)\)) underpins the low-rank nature, projecting errors into a low-dimensional boundary subspace.

To leverage this structure for solving (\ref{eqn:error_linear_sys}), we express the error term \(\alpha \mathbf{\Lambda}\) in low-rank matrix form. Using $g_l$ and $v_l$ to denote the indices and values of non-zero elements in the vector $\{\delta_{l, ijk}\}$ for each component $l$ (with length $m_l = n \cdot (n + n - 1) = 2n^2 - n$), we define the selection matrix $\mathbf{Q}_l \in \mathbb{R}^{n^3 \times m_l}$ as:
{\small
\begin{equation*}
q_{ij} = \begin{cases}
1 & \text{if } g_{l,i} = i, \, 1 \leq i \leq n^3, \, 1 \leq j \leq m_l \\
0 & \text{otherwise}
\end{cases}
\end{equation*}
\begin{equation*}
\begin{aligned}
g_l &= \{ m \mid \delta_{l,m} \neq 0, \, 1 \leq m \leq n^3 \}, \\
v_l &= \{ \delta_{l,m} \mid \delta_{l,m} \neq 0, \, 1 \leq m \leq n^3 \}
\end{aligned}
\end{equation*}
}
The block-diagonal $\mathbf{Q} = \text{diag}(\mathbf{Q}_x, \mathbf{Q}_y, \mathbf{Q}_z) \in \mathbb{R}^{3 n^3 \times (6n^2 - 3n)}$ acts as a projection operator onto the boundary subspace, while $\mathbf{W} = \text{diag}(\alpha v_x, \alpha v_y, \alpha v_z) \in \mathbb{R}^{(6n^2 - 3n) \times (6n^2 - 3n)}$ is a diagonal matrix weighting the boundary corrections (with values $\alpha \times 1$ or $\alpha \times 2$). It follows that \(\alpha \mathbf{\Lambda} = \mathbf{Q} \mathbf{W} \mathbf{Q}^T\), a low-rank outer product (rank at most $6n^2 - 3n$), as illustrated in the bottom decomposition flow of Figure \ref{fig:low_rank}(b), where the tall-skinny $\mathbf{Q}$ (purple) compresses the full system to the small $\mathbf{W}$ (orange), yielding a compact low-rank result (green).

Motivated by the need to handle this low-rank update efficiently—avoiding the prohibitive cost of reinverting the entire $3n^3 \times 3n^3$ matrix, which would lead to memory exhaustion—we apply the Woodbury formula \cite{hager1989updating}, which relates the inverse of a matrix after a low-rank perturbation to the inverse of the original matrix, transforming the large matrix inverse into a small matrix inverse plus matrix-vector multiplications:
{\small
\begin{gather*}
(\mathbf{I} + \alpha \mathbf{M} + \alpha \mathbf{\Lambda})^{-1} = \left( \mathbf{I} + \alpha \mathbf{M} + \mathbf{Q} \mathbf{W} \mathbf{Q}^T \right)^{-1} \\
= (\mathbf{I} + \alpha \mathbf{M})^{-1} - (\mathbf{I} + \alpha \mathbf{M})^{-1} \mathbf{Q} \, \mathbf{C}^{-1} \, \mathbf{Q}^T (\mathbf{I} + \alpha \mathbf{M})^{-1}
\end{gather*}
}\noindent
where the correction matrix
{\small
\[
\mathbf{C} = \left[ \mathbf{W}^{-1} + \mathbf{Q}^T (\mathbf{I} + \alpha \mathbf{M})^{-1} \mathbf{Q} \right] \in \mathbb{R}^{(6n^2 - 3n) \times (6n^2 - 3n)}
\]
}

In practice, this manifests in Algorithm \ref{alg:subdomain_solving}'s boundary correction steps, avoiding reprocessing the entire system. Without it, direct inversion would lead to prohibitive overheads, memory exhaustion, and scalability issues. Thus, we obtain the complete algorithmic procedure for solving (\ref{eqn:error_linear_sys}), presented in Algorithm \ref{alg:subdomain_solving}.

\subsection{Computational and Space Complexity Analysis}

Compared to traditional direct methods, such as LU, QR decomposition, or Gaussian elimination \cite{saad2003iterative}, which compute an explicit inverse \(\mathbf{A}^{-1}\), FlashMP reduces both computational and space complexities from \(O(n^6)\) to \(O(n^4)\), significantly reducing computational and memory overhead. For fairness, we assume the inverse matrix \(\mathbf{A}^{-1}\) is precomputed for direct methods, with runtime dominated by a general matrix-vector multiplication (GEMV). Computational complexity is measured as the count of floating-point arithmetic operations (G), while memory usage is quantified in gigabytes (GB).

\begin{table}[h]
\centering
\caption{Comparison of runtime for FlashMP and direct methods with a subdomain size of \(32^3\).}
\label{tab:complexity_flashmp}
\begin{tabular}{lcc}
\hline
\textbf{Metric} & \textbf{FlashMP} & \textbf{Direct Methods} \\
\hline
Arithmetic Operations (G) & 0.15 & 19.3 \\
Memory Usage (GB) & 0.24 & 77.3 \\
Runtime (ms) & 1.11 & 120.78 \\
\hline
\end{tabular}
\end{table}

\subsubsection{Theoretical Computational Complexity}
\paragraph{FlashMP}
The computational complexity of FlashMP is dominated by three components:
\begin{itemize}
    \item \textbf{Component-wise transforms with \(\mathbf{G}\) or \(\mathbf{G}^{-1}\)}: Each tensor product operation costs \(2n^4\) floating-point operations. The \(\mathbf{G}\) operation involves three components with three directional transforms, costing \(9 \times 2n^4 = 18n^4\) operations. The \(\mathbf{G}^{-1}\) operation is identical. Each \texttt{exact\_solve} requires \(18n^4 + 18n^4 = 36n^4\) operations, and the two \texttt{exact\_solve} calls in Algorithm \ref{alg:subdomain_solving} total \(72n^4\) operations.
    \item \textbf{Point-wise field solving}: A single block sparse matrix-vector multiplication (BSpMV) with the precomputed diagonal matrix \(\text{diag} \left( \mathbf{B}^{-1}_{ijk} \right)\), containing \(n^3\) blocks of size \(3 \times 3\), costs \(18n^3\) floating-point operations.
    \item \textbf{Boundary error correction}: Using a precomputed matrix \(\mathbf{C}^{-1} \in \mathbb{R}^{(6n^2 - 3n) \times (6n^2 - 3n)}\), matrix-vector multiplication costs \(2 \times (6n^2 - 3n)^2 \approx 72n^4\) operations.
\end{itemize}
Total complexity: \(72n^4 + 18n^3 + 72n^4 \approx O(n^4)\) operations.

\paragraph{Direct Methods}
For \(\mathbf{A^{-1}} \in \mathbb{R}^{3n^3 \times 3n^3}\), the runtime GEMV with a precomputed \(\mathbf{A}^{-1}\) costs \(2 \times (3n^3)^2 = 18n^6\) floating-point operations. 

Total complexity: \(O(n^6)\) operations.

\subsubsection{Theoretical Space Complexity}
\paragraph{FlashMP}
Memory usage includes:
\begin{itemize}
    \item Correction matrix \(\mathbf{C}^{-1} \in \mathbb{R}^{(6n^2 - 3n) \times (6n^2 - 3n)}\): \((6n^2 - 3n)^2 \times 8 \approx 288n^4\) bytes.
    \item Vectors \(\mathbf{E}, \mathbf{R} \in \mathbb{R}^{3n^3}\): \(3n^3 \times 8 = 24n^3\) bytes each.
    \item Matrices \(\mathbf{U}, \mathbf{V}\) and diagonal \(\mathbf{S} \in \mathbb{R}^{n \times n}\) : \(O(n^2)\) bytes.
\end{itemize}
Total complexity: \(O(n^4)\) bytes, dominated by \(\mathbf{C}^{-1}\).

\paragraph{Direct Methods}
Memory usage includes:
\begin{itemize}
    \item Precomputed \(\mathbf{A}^{-1} \in \mathbb{R}^{3n^3 \times 3n^3}\): \((3n^3)^2 \times 8 = 72n^6\) bytes.
    \item Vectors \(\mathbf{E}, \mathbf{R} \in \mathbb{R}^{3n^3\times 1}\): \(3n^3 \times 8 = 24n^3\) bytes each.
\end{itemize}
Total complexity: \(O(n^6)\) bytes, dominated by \(\mathbf{A}^{-1}\).

\begin{figure}[h]
  \centering
  \includegraphics[width=0.50\textwidth]{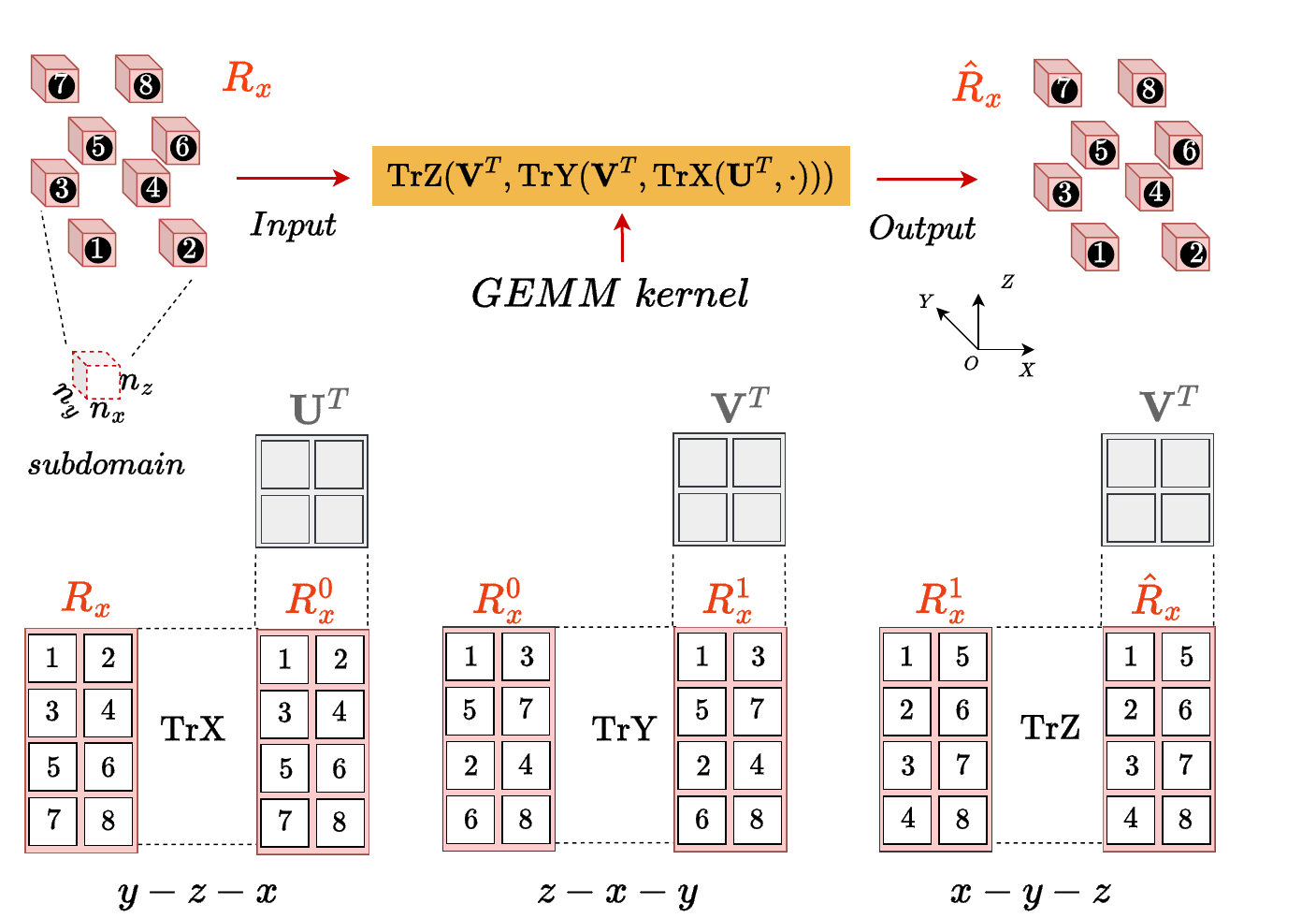}
  \caption{Tensor product operations of a field component \(R_x \) along the \(x\), \(y\), and \(z\) directions based on DGEMM. }
  \label{fig:gemm}
\end{figure}

\subsection{System Innovations}

We have also made the following  optimizations for the GPU to ensure efficient mapping of the algorithm to the hardware.

\textbf{GEMM-Based Discrete Transform.} Here the field component \(R_x \in \mathbb{R}^{n^3 \times 1}\) is a 3D tensor in the spatial domain, reshaped from a tensor of size \(n_x \times n_y \times n_z\). The tensor product operations performed by \(\text{TrX}\), \(\text{TrY}\), and \(\text{TrZ}\) in (\ref{eqn:tensor}) respectively correspond to contractions in the \(x\), \(y\), and \(z\) directions. We invoke the double-precision general matrix multiply (GEMM) interface to transpose \(R_x\) into a \(y-z-x\) format and perform the \(x\)-direction contraction with matrix parameters \(M = n_x\), \(N = n_y \times n_z\), \(K = n_x\) (for matrices of size \(M \times K\) and \(K \times N\)). The contractions in the \(y\)- and \(z\)-directions are performed similarly, with appropriate transpositions and GEMM invocations. Figure \ref{fig:gemm} provides an example of Step 1 in the FlashMP algorithm, showcasing the transformation of field variables on a \(2 \times 2 \times 2\) subdomain. Using a GEMM-based approach enables efficient, GPU-accelerated tensor operations.

\begin{figure}[h]
  \centering
  \includegraphics[width=0.50\textwidth]{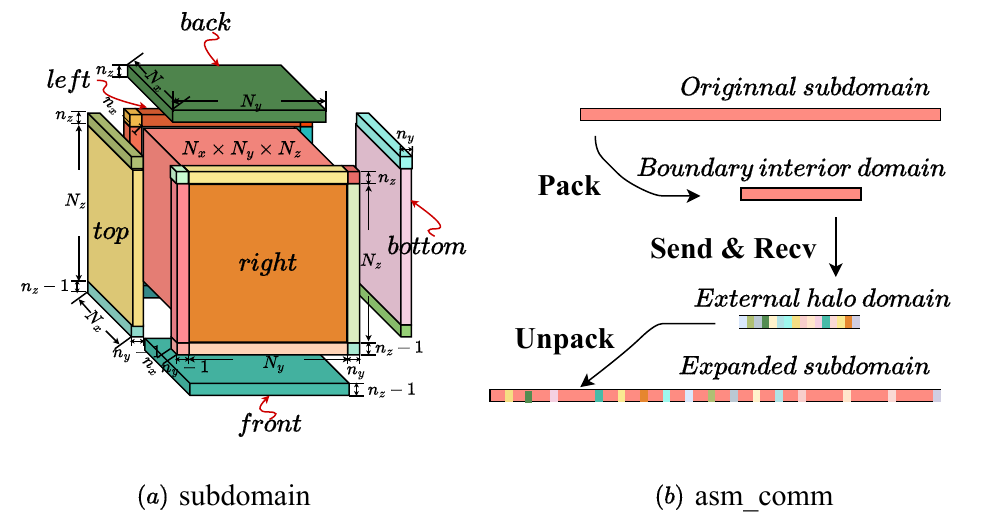}
  \caption{Inter-subdomain communication. (a) Data derived from 26 adjacent subdomains are distinguished by different colors. The block in the middle represents the data that the subdomain originally had. (b) \texttt{asm\_comm} represents the communication process involved in domain decomposition, including the three steps: Pack, Send \& Recv, and Unpack. }
  \label{fig:round}
\end{figure}

\textbf{Communication Mechanism in ASM.} Figure \ref{fig:round} illustrates the communication mechanism of the Additive Schwarz Method (ASM). As shown in Figure  \ref{fig:round}(a), each subdomain comprises an original interior domain and an extended halo domain, with the interior block representing local data and the halo capturing overlapping data. Figure  \ref{fig:round}(b) depicts the \texttt{asm\_comm} workflow, consisting of three steps. In the \emph{Pack} step, the GPU kernel extracts overlapping elements from the global vector into a send buffer. The \emph{Send \& Recv}  employs MPI non-blocking calls \cite{1996A} to exchange halo data between subdomains. Finally, the \emph{Unpack} kernel merges local and received data into an extended vector for preconditioning.

This mechanism offers two advantages. The GPU-accelerated packing and unpacking kernels enable efficient data preparation, while consolidating scattered communication into coarse-grained transfers reduces the number of MPI calls.

\section{Evaluation}
 
\subsection{Experimental Setup}

\textbf{Platform.} The test platform is an AMD GPU cluster with a Hygon C86 7185 CPU (32-core), four AMD MI60 GPUs (16 GB each), and 128 GB host memory per node. The GPUs and CPU are interconnected via PCIe, and nodes are connected via a 200 Gb/s FatTree network. The system uses ROCm 4.0 \cite{AMD_ROCm_Software} and CentOS 7.6. The GPUs have a peak performance of 5.4 TFLOPS for FP64 and a memory bandwidth of 672 GB/s.

\textbf{System setup.} 
FlashMP, as a preconditioner, can accelerate iterative solvers. To verify its effectiveness, we pair FlashMP with two representative iterative solvers: BiCGSTAB and GMRES. The efficient implementations of BiCGSTAB and GMRES on AMD GPUs are based on Hypre~\cite{falgout2002hypre}, a state-of-the-art library that provides highly optimized solver implementations for AMD GPUs.

\textbf{Workloads.} 
\texttt{GMRES} is configured with a restart length of $k =30 $, and \texttt{BiCGSTAB} uses standard parameters. Both solvers use a right-hand side vector \(\mathbf{b}\) computed as \(A\mathbf{x}_0\), where \(\mathbf{x}_0\) is a vector of random values. The stopping criterion requires a 12-order-of-magnitude reduction in the relative residual, i.e., \(\frac{\|\mathbf{b} - A\mathbf{x}_k\|}{\|\mathbf{b}- A\mathbf{x}_0\|} < 10^{-12}\). Experiments test a fixed subdomain size of \(32^3\) per GPU, without preconditioning and with FlashMP at overlap 0 to 3.

\textbf{Metrics.} 
The total time \( T_{\text{total}} \) of the iterative solver is:
\begin{equation*}
T_{\text{total}} = \#\text{iter} \cdot T_{\text{single}}
\end{equation*}
where \(\#\text{iter}\) is the number of iterations, and \( T_{\text{single}} \) is the single-iteration time, decomposed into precond time \( T_{\text{precond}} \) and core subspace time \( T_{\text{core}} \). FlashMP minimizes \(\#\text{iter}\) through effective exact preconditioning and optimizes \( T_{\text{single}} \) via efficient implementation. Combining the two sides leads to an overall high performance solver. 


\begin{figure}[h]
\centering
\begin{subfigure}[t]{0.85\linewidth}
    \centering
    \includegraphics[width=\linewidth]{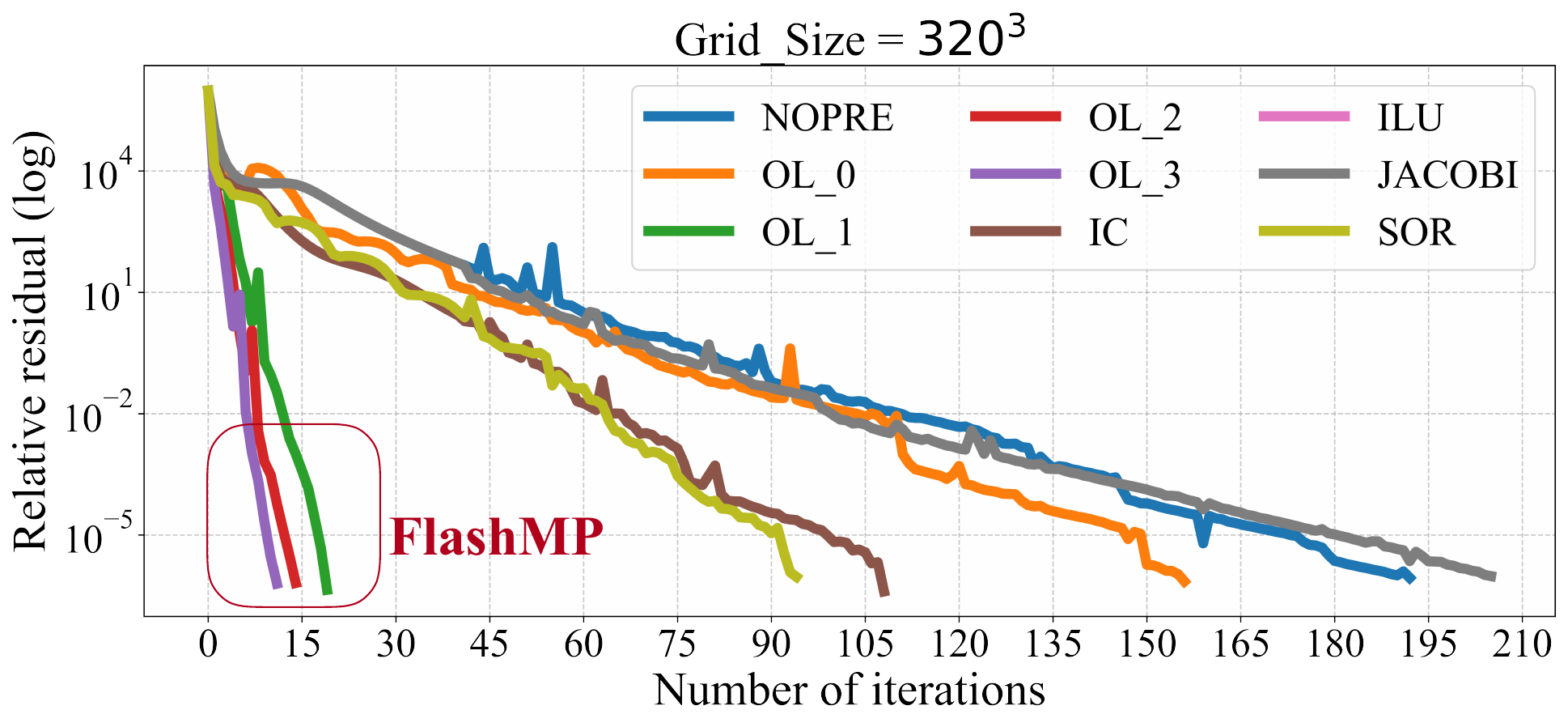}
    \caption{Convergence traces for \texttt{BiCGSTAB}.}
    \label{fig:BiCGSTAB-residual}
\end{subfigure}
\vspace{0.5cm}
\begin{subfigure}[t]{0.85\linewidth}
    \centering
    \includegraphics[width=\linewidth]{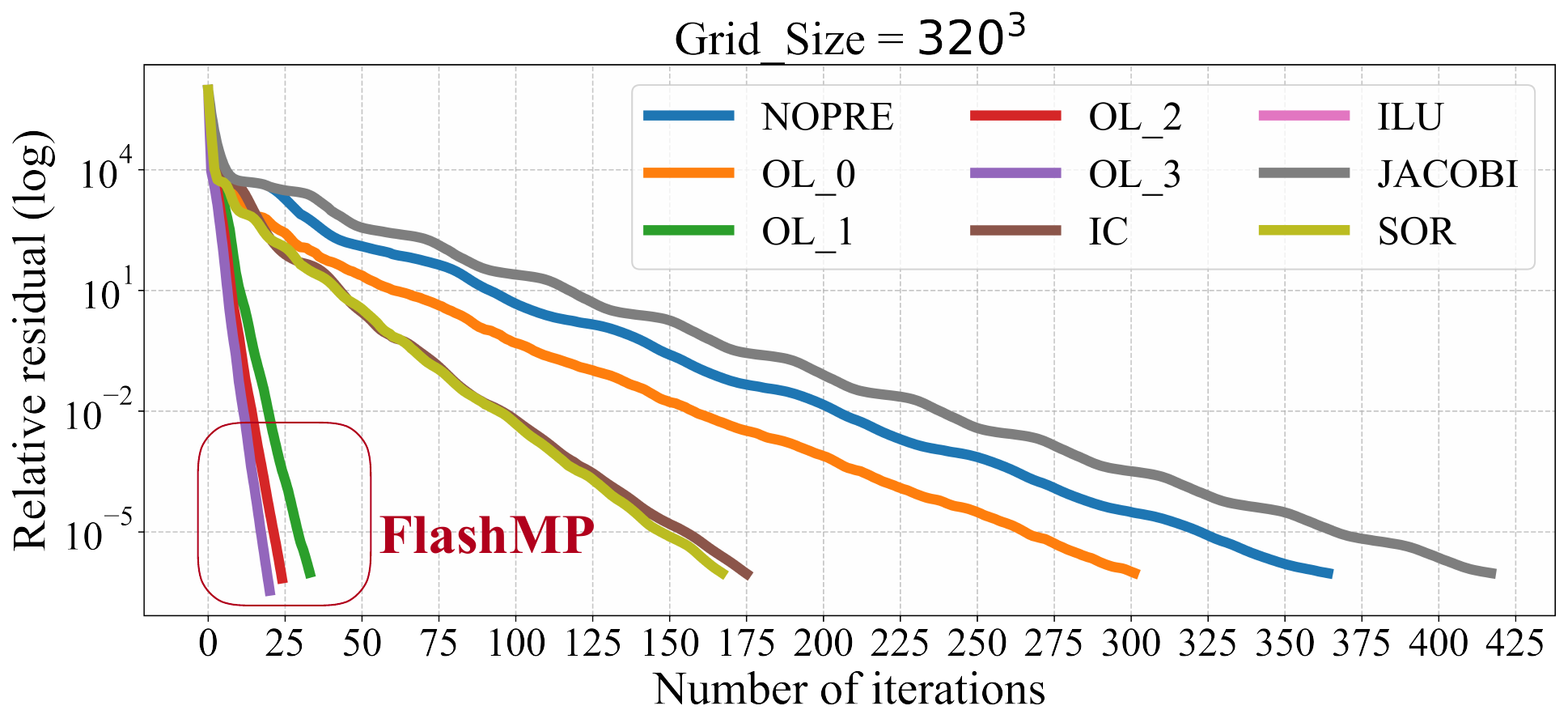}
    \caption{Convergence traces for \texttt{GMRES}.}
    \label{fig:GMRES-residual}
\end{subfigure}
\caption{Convergence curves of \texttt{BiCGSTAB} (a) and \texttt{GMRES} (b) with different preconditioners, where ``\texttt{NOPRE}'' represents without preconditioner, ``\texttt{OL\_i}'' represents the use of the FlashMP with overlap $i$, and ``\texttt{ILU}'', ``\texttt{IC}'', ``\texttt{SOR}'' represent incomplete LU, incomplete Cholesky, and successive over-relaxation, respectively. The Y axis is the relative residual, and the X axis is the iteration number.}
\label{fig:convergence-traces}
\end{figure}

\subsection{Convergence Analysis}

Figure \ref{fig:convergence-traces} shows convergence curves of iterative solvers with various preconditioners. Unlike approximate methods (ILU, ICC, Jacobi, SOR), FlashMP’s subdomain exact solving achieves the fastest convergence with significantly lower single-iteration time than traditional exact direct methods (in table \ref{tab:complexity_flashmp}). Existing approximate methods on GPU fail to reduce iterations substantially while increasing single-iteration overhead, rendering them uncompetitive with NOPRE in total time; thus, we focus on comparing FlashMP and NOPRE.

For \texttt{BiCGSTAB} in Figure \ref{fig:convergence-traces}(a), the NOPRE case converges in 193 iterations, while FlashMP with overlap 0, 1, 2, and 3 converges in 157, 20, 15, and 12 iterations, respectively. The residual curve of \texttt{BiCGSTAB} shows oscillations, indicating numerical instability in its biconjugate framework. Overlap 0 offers a little reduction in iterations due to limited subdomain data exchange, but overlap 1 reduces iterations significantly to 20, with minor further gains at overlap 2 and 3. For \texttt{GMRES} in Figure \ref{fig:convergence-traces}(b), the NOPRE case requires 364 iterations, while FlashMP with overlap 0, 1, 2, and 3 converges in 301, 33, 24, and 20 iterations, respectively. The residual curve of GMRES exhibits  smoother, which is attributed to stability from orthogonalization. Similar to \texttt{BiCGSTAB}, overlap 0 provides minor benefits, highlighting the importance of adequate overlap. \texttt{BiCGSTAB} converges about twice as fast as \texttt{GMRES} due to its dual preconditioning per iteration.

\begin{figure}[h!]
  \centering
  \begin{subfigure}{\linewidth}
    \centering
    \includegraphics[width=0.95\linewidth]{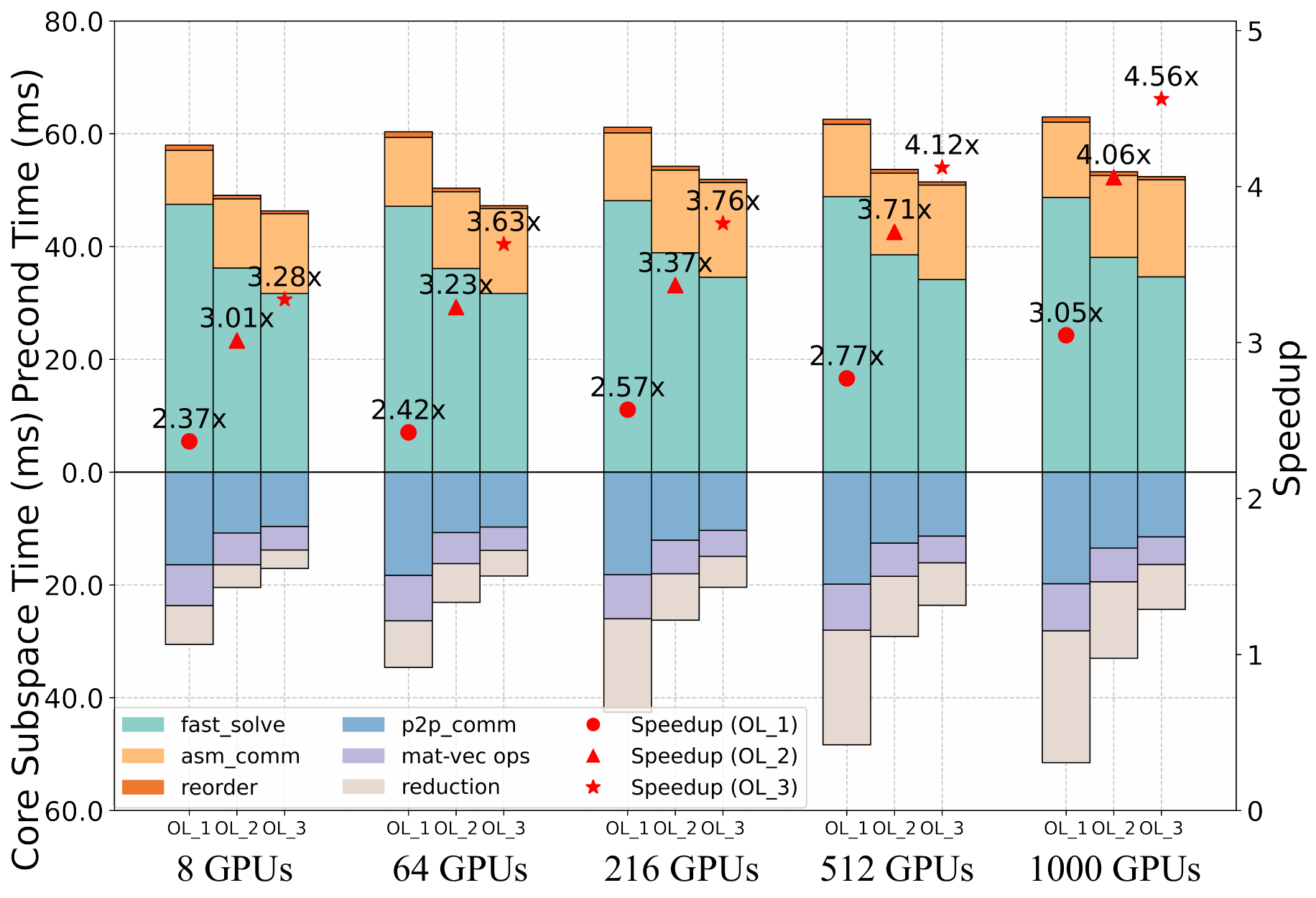}
    \caption{Breakdown for \texttt{BiCGSTAB}.}
    \label{fig:BiCGSTAB_breakdown}
  \end{subfigure}
  \vskip 0.2cm
  \begin{subfigure}{\linewidth}
    \centering
    \includegraphics[width=0.95\linewidth]{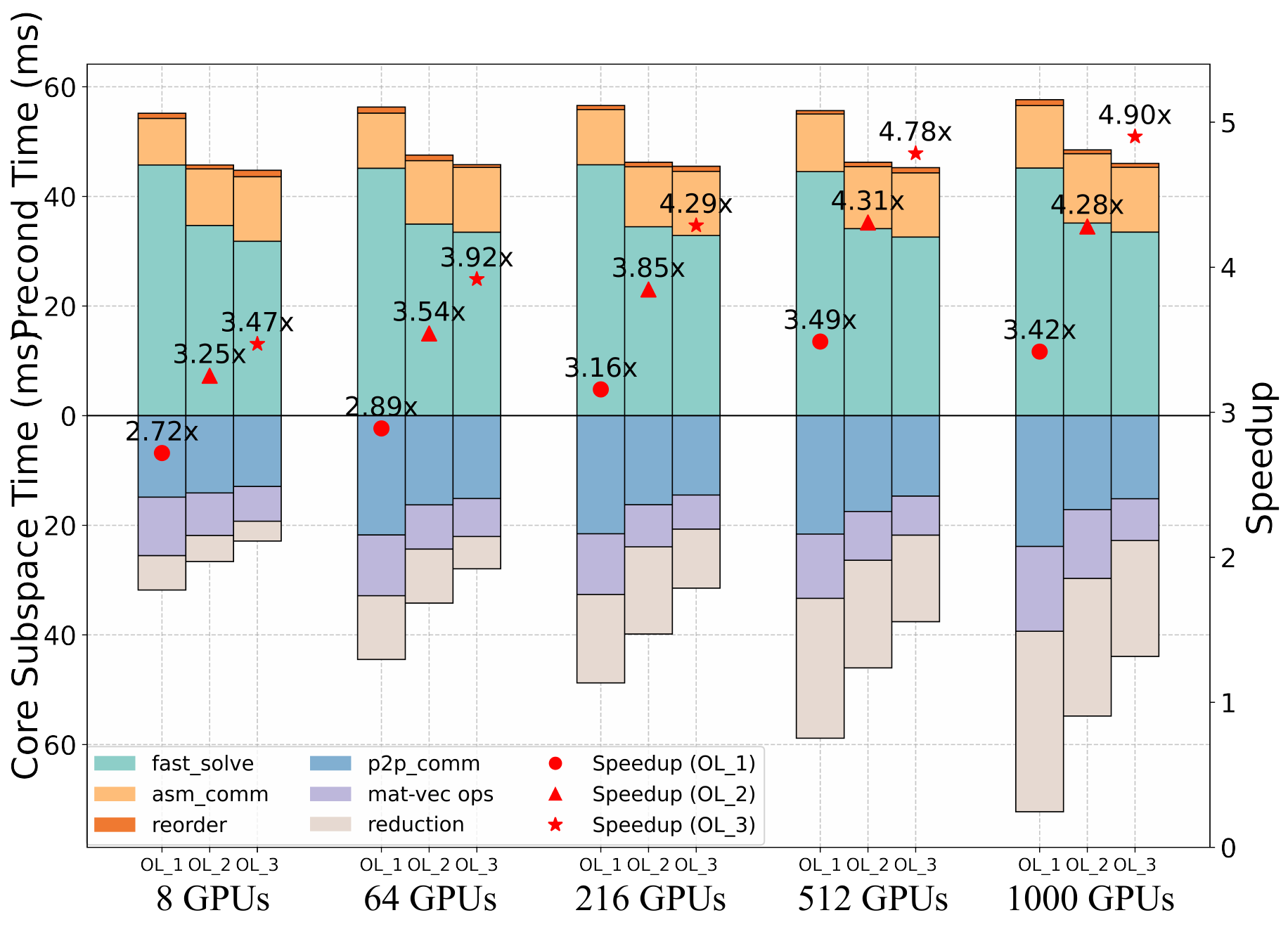}
    \caption{Breakdown for \texttt{GMRES}.}
    \label{fig:GMRES_breakdown}
  \end{subfigure}
  \caption{Time breakdown for \texttt{BiCGSTAB} (a) and \texttt{GMRES} (b) into precond time \( T_{\text{precond}} \) (reorder, asm\_comm, fast\_solve) and core subspace time \( T_{\text{core}} \) (point-to-point communication, matrix-vector operations, reduction).  {fast\_solve}  represents the time taken by FlashMP to perform exact subdomain solves. Speedup is relative to the NOPRE across various GPU counts with overlap 1, 2, and 3. Times are in milliseconds.}
  \label{fig:breakdown_comparison}
\end{figure}

\subsection{Performance and Speedup}

Figure \ref{fig:breakdown_comparison} quantifies the speedup of FlashMP paired with \texttt{BiCGSTAB} and \texttt{GMRES}, decomposing total time into precond time  \( T_{\text{precond}} \) and core subsparse time \( T_{\text{core}} \) across GPU counts of 8, 64, 216, 512, and 1000. Overlap 0 is excluded because it does not significantly reduce iteration counts while increasing single-iteration time \( T_{\text{single}} \), making it uncompetitive with the NOPRE case.  

For \texttt{BiCGSTAB} in Figure \ref{fig:breakdown_comparison}(a) at 1000 GPUs, NOPRE runtime is 363.8~ms, while FlashMP yields 119.4~ms, 89.6~ms, and 79.8~ms for overlaps 1, 2, and 3, achieving speedups of \(3.05\times\), \(4.06\times\), and \(4.56\times\), respectively, due to fewer iterations (20, 15, 12 vs. 193). For \texttt{GMRES} in Figure \ref{fig:breakdown_comparison}(b), NOPRE runtime is 440.7~ms, with FlashMP times of 89.9~ms, 103.3~ms, and 129.9~ms for overlaps 1, 2, and 3, yielding speedups of \(3.42\times\), \(4.28\times\), and \(4.90\times\), driven by iteration reductions (33, 24, 20 vs. 364).

\begin{table}[h!]
\centering
\caption{Work requirements per iteration for \texttt{BiCGSTAB} and \texttt{GMRES}. Here, \( k_{\text{avg}} \), the average number of orthogonal vectors per \texttt{GMRES} iteration, is approximately \( \frac{k + 1}{2} \).}
\begin{tabular}{lcc}
\toprule
Operation Type & \texttt{BiCGSTAB} & \texttt{GMRES} \\ \midrule
Preconditoning & 2 & 1 \\
Matrix-vector multiplication (SpMV) & 2 & 1 \\
Dot product & 4 & \( k_{\text{avg}} + 1 \) \\
Scalar-vector multiplication (AXPY) & 6 & \( k_{\text{avg}} \) \\ \bottomrule
\end{tabular}
\label{tab:work_requirements}
\end{table}

For both solvers, increasing overlap reduces total time, as shown in Figure \ref{fig:breakdown_comparison}, by strengthening the preconditioner and lowering iteration counts. Although single-iteration time rises with larger subdomains, the substantial iteration reduction outweighs this cost. \textit{fast\_solve} dominates precond time, while reduction time, driven by collective communication, grows logarithmically with GPU count, increasing its share. FlashMP’s exact preconditioning minimizes iterations, enhancing speedup at higher GPU counts by reducing synchronization overhead, especially evident at scale. Table \ref{tab:work_requirements} shows \texttt{BiCGSTAB} requires one additional precond step and SpMV \texttt{GMRES}, reflecting its bi-orthogonalization approach with dual updates. Thus, \texttt{BiCGSTAB} needs at least twice the residual norm decrease per iteration to compete with \texttt{GMRES}.

\begin{figure}[h]
  \centering
  \begin{subfigure}{0.42\textwidth}
    \centering
    \includegraphics[width=\linewidth]{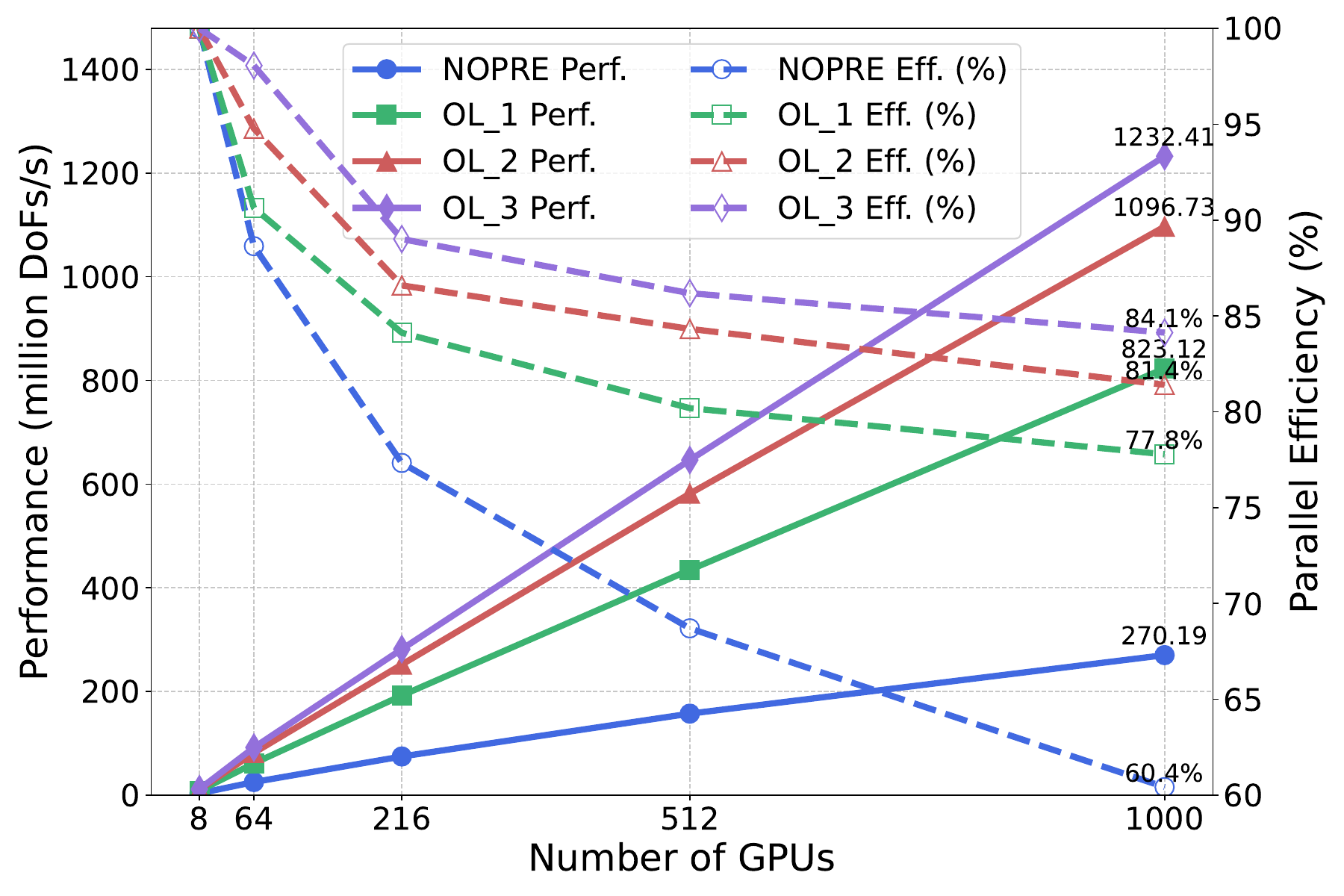}
    \caption{Weak scalability for \texttt{BiCGSTAB}.}
    \label{subfig:BiCGSTAB_scaling}
  \end{subfigure}
  \vskip 0.2cm
  \begin{subfigure}{0.42\textwidth}
    \centering
    \includegraphics[width=\linewidth]{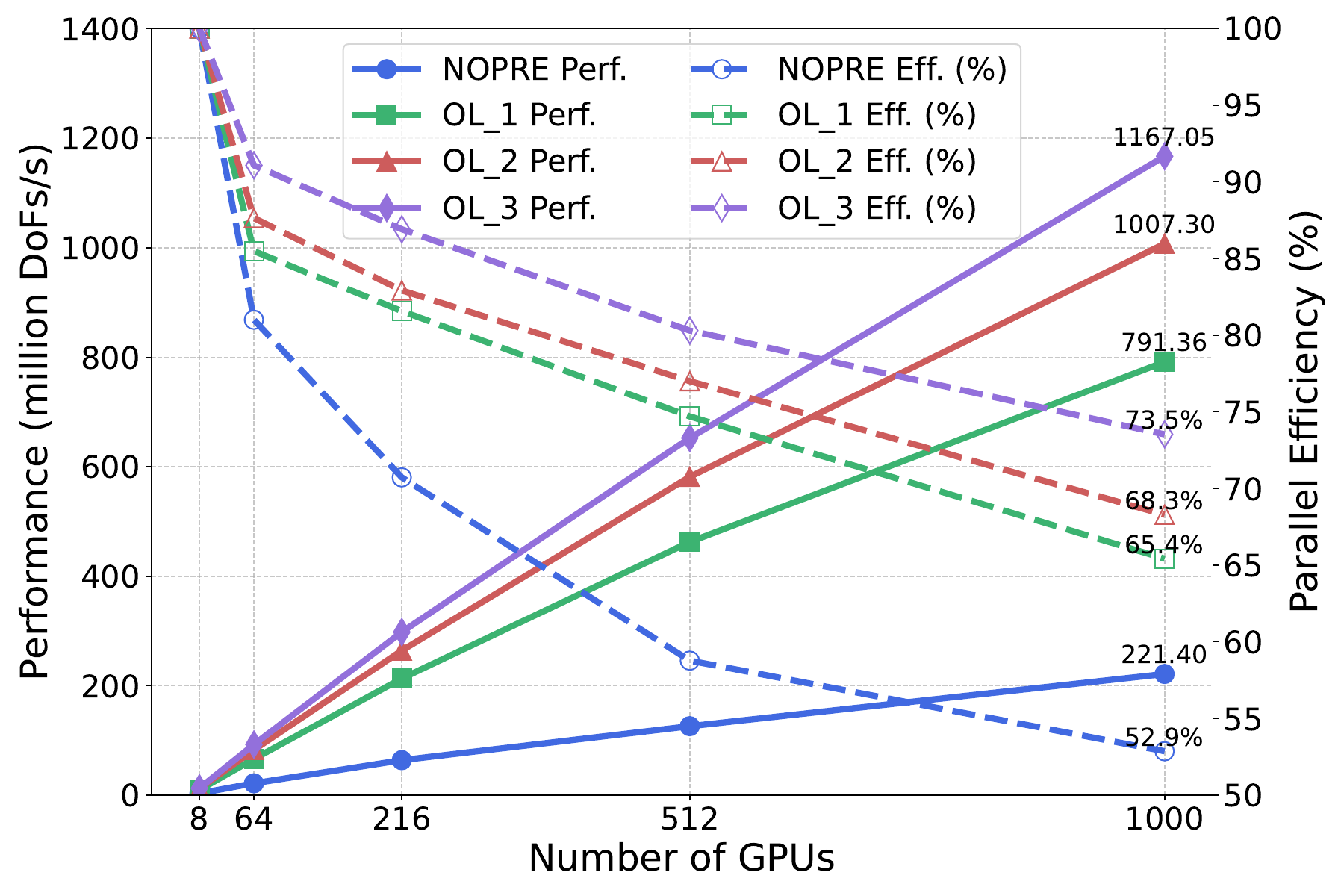}
    \caption{Weak scalability for \texttt{GMRES}.}
    \label{subfig:GMRES_scaling}
  \end{subfigure}
  \caption{Weak scalability for \texttt{BiCGSTAB} (a) and \texttt{GMRES} (b). Solid lines represent performance on the left y-axis, and dashed lines represent parallel efficiency on the right y-axis.}
  \label{fig:scaling_comparison}
\end{figure}

\subsection{Parallel Scalability} 

Parallel efficiency was evaluated by scaling GPUs from 8 to 1000, maintaining a fixed subdomain size of $32^3$ per GPU. Performance is measured in million degrees of freedom per second (MDoF/s), with parallel efficiency relative to ideal linear scaling from 8 GPUs. 

For \texttt{BiCGSTAB} in Figure \ref{fig:scaling_comparison}(a), the NOPRE achieves 262.96 MDoF/s with 63.4\% efficiency at 1000 GPUs. FlashMP with overlap 1, 2, and 3 reach 823.12, 1096.73, and 1232.41 MDoF/s, with efficiencies of 77.8\%, 81.4\%, and 84.1\%, respectively. For \texttt{GMRES} in Figure \ref{fig:scaling_comparison}(b), the NOPRE reaches 221.40 MDoF/s with 52.9\% efficiency at 1000 GPUs. FlashMP with overlap 1, 2, and 3 yields 791.36, 1007.30, and 1167.05 MDoF/s, with efficiencies of 65.4\%, 68.3\%, and 73.5\%, respectively. \texttt{BiCGSTAB} outperforms \texttt{GMRES} in scalability, achieving up to 84.1\% efficiency versus \texttt{GMRES}’s 73.5\% at 1000 GPUs, driven by its fixed communication costs. In large-scale parallel contexts, communication overhead is a key factor affecting performance. The orthogonalization of \texttt{GMRES} requires dot products and reductions across all orthogonal vectors, which demand more collective communication and global synchronization compared to \texttt{BiCGSTAB}. This can lead to a decrease in parallel efficiency. The fixed number of operations in \texttt{BiCGSTAB} makes it more suitable for large-scale parallelism, see table \ref{tab:work_requirements}. Scalability tests demonstrate FlashMP’s effectiveness in enhancing parallel efficiency, as its subdomain exact solving ensures minimal iteration counts, effectively reducing synchronization counts in global communication---a critical factor limiting parallel efficiency.

\section{Conclusions}
This paper introduces FlashMP, a high-performance preconditioner tailored for solving large-scale linear systems in electromagnetic simulations. FlashMP effectively decouples the ill-conditioned double-curl operator by combining domain decomposition and discrete transforms, significantly reducing iteration counts and computational overhead across various conditions. Extensive testing on distributed GPU clusters reveals that FlashMP decreases iteration counts by up to 16$\times$ and achieves speedups ranging from 2.5$\times$ to 4.9$\times$ over NOPRE implementation in state-of-the-art libraries Hypre. Furthermore, weak scalability tests demonstrate parallel efficiencies up to 84.1\%  at 1000 GPUs.


\section*{Acknowledgment}
The authors express their gratitude to the anonymous reviewers for their insightful comments and constructive suggestions. This work is supported by the Strategic Priority Research Program of Chinese Academy of Sciences, Grant NO.XDB0500101. Jian Zhang is the corresponding author of this paper (\href{mailto:zhangjian@sccas.cn}{zhangjian@sccas.cn}).

\bibliographystyle{plain}
\bibliography{reference.bib}

\end{document}